\documentclass[conference]{IEEEtran}
\IEEEoverridecommandlockouts
\usepackage[left=0.625in,right=0.625in,top=0.675in,bottom=1in]{geometry}
\usepackage{cite}
\usepackage{amsmath,amssymb,amsfonts}
\usepackage{algorithmic}
\usepackage[linesnumbered,ruled]{algorithm2e}
\usepackage{graphicx}
\usepackage{textcomp}
\usepackage{xcolor}

\usepackage[linesnumbered,ruled]{algorithm2e}

\usepackage{graphics,epstopdf}
\usepackage{balance}
\usepackage{listings}
\usepackage{comment}
\usepackage{enumitem}
\usepackage{fixmath}
\usepackage{amsmath,amssymb,amsfonts}
\usepackage{tabularx}
\usepackage{booktabs}
\usepackage{longtable}
\usepackage{multirow}
\usepackage{colortbl}
\usepackage{stmaryrd}
\usepackage{enumitem}
\usepackage{soul}
\usepackage{bm} 
\usepackage{mathtools} 
\usepackage{mathrsfs} 
\usepackage{multirow}
\usepackage{array}
\usepackage{hhline, boldline, makecell}

\usepackage{cellspace}
\def\BibTeX{{\rm B\kern-.05em{\sc i\kern-.025em b}\kern-.08em
    T\kern-.1667em\lower.7ex\hbox{E}\kern-.125emX}}
\begin{document}

\title{ Support Vector Machine Implementation on MPI-CUDA and Tensorflow Framework}

\author{
Islam Elgarhy   
\\
Department of Computer Systems, Faculty of Computer and Information Sciences,\\ Ain Shams University, Cairo, Egypt.\\
islam$\_$elgarhi@cis.asu.edu.eg
}

\maketitle

\begin{abstract}


Support Vector Machine (SVM) algorithm requires a high computational cost (both in memory and time) to solve a complex quadratic programming (QP) optimization problem during the training process. Consequently, SVM necessitates high computing hardware capabilities. The central processing unit (CPU) clock frequency cannot be increased due to physical limitations in the miniaturization process. However, the potential of parallel multi-architecture, available in both multi-core CPUs and highly scalable GPUs, emerges as a promising solution to enhance algorithm performance. Therefore, there is an opportunity to reduce the high computational time required by SVM for solving the QP optimization problem. This paper presents a comparative study that implements the SVM algorithm on different parallel architecture frameworks. The experimental results show that SVM MPI-CUDA implementation achieves a speedup over SVM TensorFlow implementation on different datasets. Moreover, SVM TensorFlow implementation provides a cross-platform solution that can be migrated to alternative hardware components, which will reduces the development time.


\end{abstract}


\section{Introduction}
\label{sec:introduction}

Recently, machine learning (ML) has been used in many different fields of our life. ML aims to allow machines to learn as humans do. There are many different algorithms that aim to manipulate offline dataset information to extract knowledge, enabling machines to learn how to predict new information. ML algorithms aim to learn a function $f(x)=y$ that maps input data x onto output $y$. The core of any ML algorithm is to learn a function $f$ that allows it to identify and map new input $x$ to its output $y$.

Support Vector Machine (SVM) is one of the core algorithms in ML. SVM is widely used in both classification and regression processes. SVM is a binary classifier. One-to-One and One-to-Many are two approaches to implementing multiclass. SVM requires high computational time, intensive memory, and storage requirements to solve large-scale problems. However, the performance of the central processing unit (CPU) cannot be increased due to limitations of clock frequencies. Therefore, significant improvements have been made through high-scalable graphical processing unit (GPU) parallel architecture and multi-core processors. These improvements provide a chance to overcome the problems that require high computational cost \cite{flynn1972some,cao2008network,tokhi2003parallel,bekkerman2011scaling}.


Message Passing Interface (MPI) is a standardized means of communication between different nodes across distributed memory. Compute Unified Device Architecture (CUDA) from NVIDIA was developed as a parallel computing platform and programming model to leverage the power of the GPU. A hybrid model combining both MPI-CUDA can be used to best utilize the hardware capabilities \cite{bekkerman2011scaling,aoyama1999rs}. TensorFlow is an open-source machine learning library that allows application program interfaces (APIs) for implementing ML algorithms on a variety of platforms (CPU and GPU) for desktop, mobile, web, and cloud \cite{abadi2016tensorflow}.


The rest of this paper is organized as follow. Section \ref{sec:sec2} introduces the hybrid message passing interface - compute unified device architecture (MPI-CUDA) and Tensorflow framework. The introduction about SVM algorithm and our proposed technique design is presented and discussed in section \ref{sec:sec3}. Finally, the experimental results and datasets are described in section \ref{sec:sec4}, and the conclusion is provided in section \ref{sec:sec5}.

\section{Parallel Architecture}
\label{sec:sec2}

\subsection{Message Passing Interface - Compute United Device Architecture (MPI-CUDA)}

MPI is a standard interface for communication between different nodes in distributed memory, where each node has its local address space and runs independently. MPI provides node-to-node communication and collective communication through the communication network between the nodes. The MPI programming model follows the Single Program Multiple Data (SPMD) model, where a set of nodes executes the same instructions on different data. The instructions are loaded onto all nodes, but each node can follow a different execution path. There are many implementations that follow the MPI standard; the most popular of them are MPICH and LAMMPI \cite{aoyama1999rs}.


CUDA is an API used for leveraging GPU parallel architecture in General-Purpose Programming (GPGPU). NVIDIA introduced the CUDA API for the first time in November 2006 to overcome the limitations of using GPUs in GPGPU due to the difficulty of knowing and interacting with DirectX or OpenGL. In the CUDA programming model, the GPU can be considered as a massive parallel Single Instruction Multiple Data (SIMD) Streaming Multiprocessor (SM) array. A group of grid/block/thread can be assigned to execute a kernel \cite{cook2012cuda}.


A kernel is a program to be executed with each thread, using thread identifier and block identifier. Each thread can perform the kernel task on a different set of data in parallel. The CPU, or 'host,' is responsible for launching the kernel to be executed on the GPU, or 'device.' It also sends input kernel data from its memory 'host memory' to the GPU 'device memory. Then, it copies the result back from 'device memory' to 'host memory' Each SM has its own shared memory, which is common to all the processors inside it. It also has constant memory caches. Communication between different SMs occurs through device memory, which is available to all the processors of the SM \cite{cook2012cuda,nvidia2010nvidia,sanders2010cuda}.

\begin{figure}[!t]
   \centering
	\includegraphics[width=\linewidth]{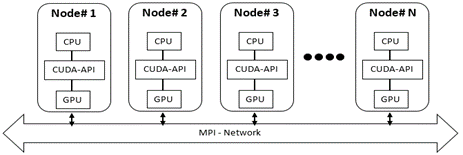}
 \caption{Hybird Memeory Archirtecture. 
 }
    \label{fig1}
     \vspace{-0.35cm}
\end{figure}

MPI represents a distributed memory architecture, while CUDA represents a shared memory architecture. The hybrid memory architecture combines both distributed and shared memory architectures. Each node has its own multicore processor that can be utilized as part of a shared architecture memory model. Different nodes are considered part of a distributed memory architecture, communicating through a network and using various communication directives to exchange messages among them \cite{khaled2015design,chen2009stream}. This is illustrated in Fig.\ref{fig1}.


\subsection{Tensorflow Framework}

Recently, many efficient and easy-to-use ML frameworks have been introduced to accommodate the increasing complexity of ML algorithms. They allow beginners and experts to develop their own ML techniques. TensorFlow is one such ML framework library. TensorFlow is an open-source ML library interface for expressing ML algorithms and executing predefined ML algorithms

\begin{figure}[!t]
   \centering
	\includegraphics[width=\linewidth]{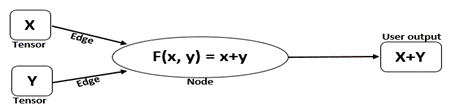}

 \caption{Tensorflow Graph Example. 
 }
    \label{fig2}
     \vspace{-0.35cm}
\end{figure}

TensorFlow is used for high-performance numerical computation. Its flexible architecture allows easy implementation and deployment on cross-platforms (desktops, clusters of servers, CPUs, GPUs, TPUs, and mobile devices), and it can run on several operating systems \cite{abadi2016tensorflow}. TensorFlow is an expression where a tensor is data, and flow is the expressing flow for this data in a dataflow graph ($TensorFlow = Tensor + Flow = Data + Flow$). A dataflow graph is a form of a directed graph consisting of a set of nodes and edges between them, where nodes describe instructions, and edges represent the data flow \cite{abadi2016tensorflow}. TensorFlow implementations are represented in two steps; the first of them is constructing a dataflow graph, the second uses a session to run instructions in this graph and evaluate the output result. The session is the runtime environment of a graph, where instructions are executed, and tensors are calculated, as shown in Fig.\ref{fig2}.


\section{Support Vector Machine algorithm and Proposed technique design }\label{sec:sec3} 

\subsection{Sequential SVM}

SVM is binary classifier ML algorithm. Given a set of training data samples $X =\{x_1, x_2 \dots x_n\}$, where n is the size of $X$ , $x \in R_d$ and d input feature size, $yi \in \{1,-1\}$ is the class label of input sample $x_i$. Assuming classes are linearly separable, SVM aim to find a hyperplane that gives the maximum margin between two classes of data in $X$. "Support vectors" are the vectors of both classes located on the margin hyperplane \cite{chang2011libsvm}. 
In most cases, training data samples are not linearly separable, so this hyperplane is not exist. Using Kernel function gives advantage to SVM for non-linearly separable classes \cite{catanzaro2008fast}.
However SVM is a binary classifier technique, but it can be used as multi-class classifier. The “one-against-one” is the more suitable method for practical use than other methods. For $m$ classes, there are $m(m-1)/2$ independent binary classification problems \cite{hsu2002comparison}.

\subsection{MPI-CUDA SVM}

Sequential minimum optimization (SMO) algorithm used to solve the SVM binary training quadratic problem, where training data samples are divided into smaller working set. SMO solves the smallest task at each step to find only two optimization variables and update two Lagrange multipliers under the Karush-Kuhn-Tucker (KKT) constraints \cite{scholkopf1999advances,fan2005working,keerthi2001improvements}.

\begin{figure}[!t]
   \centering
	\includegraphics[width=\linewidth]{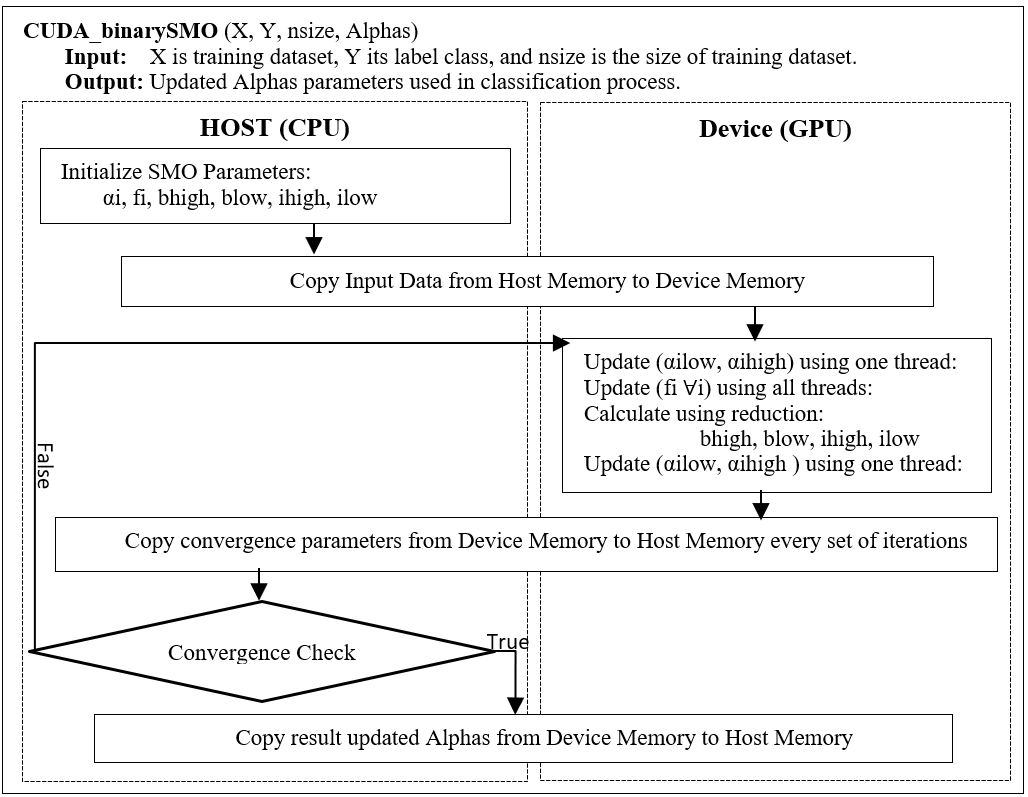}
 \caption{CUDA Binary-Class SMO. 
 }
    \label{fig3}
     \vspace{-0.35cm}
\end{figure}

\begin{figure}[!t]
   \centering
	\includegraphics[width=\linewidth]{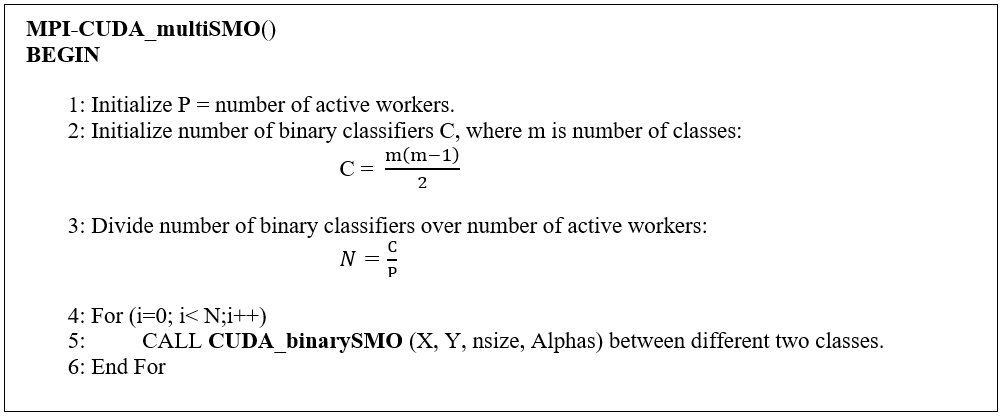}
 \caption{MPI-CUDA Multi-Class SMO. 
 }
    \label{fig4}
     \vspace{-0.35cm}
\end{figure}

In \cite{lopes2015gpu,tan2015gpu}, the SMO is implemented in parallel, making full use of the advantage of the huge computational power of GPU devices. In MPI-CUDA implementation \cite{elgarhy2019multi}, we launched a thread per independent training data sample. Hence, most of the SMO computation steps can run in parallel. Then, convergence checks were executed on the host for every set of iterations on the device, as shown in Fig.\ref{fig3}. Moreover, Fig.\ref{fig4} illustrates the use of Hybrid MPI-CUDA to execute the multi-class training problem. It involves running multiple parallel binary SMOs to implement a parallel multi-class SMO. We distribute the number of parallel binary SMOs among the MPI working nodes \cite{elgarhy2019multi}.

\subsection{Tensorflow SVM}

As with many implementations of algorithms in TensorFlow, SVM is described as a directed graph consisting of nodes and edges, where nodes represent instructions (operations), and edges represent data flow. A 'Variable' is a special instruction (operation) without computation; it is used only for storing any parameters of a described algorithm. A 'Placeholder' is a tensor that will always be fed. After graph construction, a TensorFlow session is used to execute the graph and compute the result \cite{goldsborough2016tour,yuan2017comparison}.

\begin{figure}[!t]
   \centering
	\includegraphics[width=\linewidth]{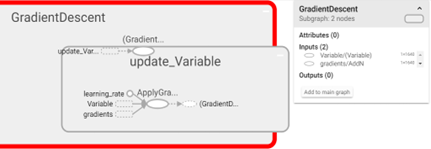}
 \caption{Tensorboard Gradient Descent Optimizer for binary-class. 
 }
    \label{fig5}
\end{figure}

For the SVM binary training problem, the directed graph consists of three steps. The first of them is initializing the input training data samples using 'Placeholders'. The second step is defining SVM 'Variables' and describing the Gaussian RBF kernel function. The third step is describing the SVM model and declaring the gradient descent optimizer algorithm, as shown in Fig.\ref{fig5}. After describing the graph for SVM binary training process, a TensorFlow session is used to run it and compute the output result. Multiple running sessions are used to implement the SVM multiclass training process with the 'one-against-one' approach.


\section{Datasets \& Experimental Results} \label{sec:sec4}

In this section, we present the datasets used to compare SVM implementations on both CUDA and Tensorflow. It highlights the performance of the CUDA implementation over the Tensorflow implementation. It also shows the cross-platform nature of Tensorflow implementation, demonstrating its ability to be migrated to alternative hardware components.


\subsection{Datasets}

Table \ref{table1}  shows the experimental datasets. The first dataset is a hyperspectral dataset consisting of 1096x715 pixels and nine ground-truth classes labeled as water, trees, grass, parking lot, bare soil, asphalt, bitumen, tiles, and shadow. The second dataset is the Iris flower multivariate dataset with 150 total data samples and 3 classes (setosa, virginica, versicolor). The third dataset is the Breast Cancer Wisconsin dataset with 569 data samples and 2 possible classes (benign, malignant).


\begin{table}[!t]
\centering
\caption{Datasets.}
\label{table1}
\resizebox{\columnwidth}{!}
{%
\renewcommand{\arraystretch}{1.2}
\begin{tabular}{|c|c|c|c|} 
\hline
\rowcolor[rgb]{0.753,0.753,0.753} \textbf{Dataset} & \textbf{Description}                                                                                                                               & \textbf{\#Classes} & \textbf{\#Features}  \\ 
\hline
Pavia
Centre                                       & \begin{tabular}[c]{@{}c@{}}PaviaCentre hyperspectral \\dataset was over Pavia city center, Italy.\end{tabular}                                     & ~9                 & 102 Spectral
bands   \\ 
\hline
Iris
Flower                                        & \begin{tabular}[c]{@{}c@{}}Iris flower multivariate dataset \\was obtained by Ronald Fisher \\the british statistician and biologist.\end{tabular} & 3                  & 4 Features           \\ 
\hline
Breast
Cancer                                      & \begin{tabular}[c]{@{}c@{}}BreastCancer Wisconsin dataset\\was obtained from the University of wisconsinhospitals.\end{tabular}                    & 2                  & 32
Features          \\
\hline
\end{tabular}
}
\end{table}

\subsection{Experimental Results}

Table \ref{table2} shows the hardware specifications for the machine used to carry out these experiments. CUDA Version 9.0 and MPICH2 are used to implement SVM MPI-CUDA, while SVM Tensorflow is implemented using TensorFlow version 1.8.0. Both implementations run on Windows 10 (64-bit). 

\begin{figure}[!t]
   \centering
	\includegraphics[width=\linewidth]{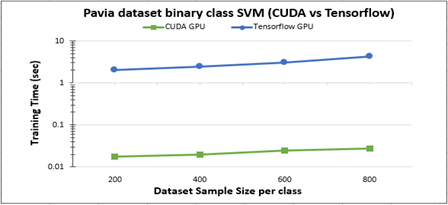}
 \caption{Binary training time CUDA vs. Tensorflow. 
 }
    \label{fig6}
\end{figure}

\begin{figure}[!t]
   \centering
	\includegraphics[width=\linewidth]{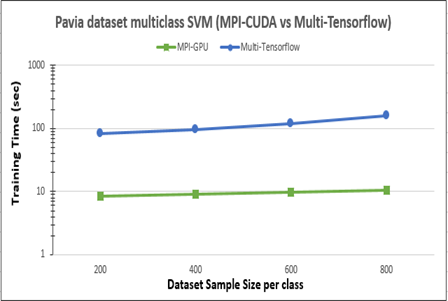}
 \caption{Multi-Class training time MPI-CUDA vs. Multi-Tensorflow. 
 }
    \label{fig7}
\end{figure}

For the Pavia Centre dataset, Table \ref{table3} and Fig.\ref{fig6} present the binary training times at various sample sizes for a binary class SVM. This comparison involves the CUDA-GPU implementation against the Tensorflow-GPU implementation. Specifically, CUDA-GPU achieved a speedup of 154.3X over Tensorflow-GPU with 800 sample points for each class. Similarly, Table \ref{table4} and Fig.\ref{fig7} present the training times for multiclass SVM on different sample sizes using the hybrid MPI-CUDA implementation over Multi-Tensorflow for the same dataset. MPI-CUDA achieved a speedup of 14.9X over Multi-Tensorflow with 800 sample points per each class.


\begin{table}[!t]
\centering
\caption{Machine Hardware Specs (CPU-GPU).}
\label{table2}

\centering
{%
\renewcommand{\arraystretch}{1.2}
\begin{tabular}{|c|c|} 
\hline
{\cellcolor[rgb]{0.753,0.753,0.753}}                                        & CPU Core i7-7500M @2.70 GHz ,2.9 GHz  \\ 
\hhline{|>{\arrayrulecolor[rgb]{0.753,0.753,0.753}}->{\arrayrulecolor{black}}-|}
\multirow{-2}{*}{{\cellcolor[rgb]{0.753,0.753,0.753}}\textbf{Host(CPU) }}   & 16 GB RAM                             \\ 
\hline
{\cellcolor[rgb]{0.753,0.753,0.753}}                                        & NVIDIA GeForce GTX950M                \\ 
\hhline{|>{\arrayrulecolor[rgb]{0.753,0.753,0.753}}->{\arrayrulecolor{black}}-|}
\multirow{-2}{*}{{\cellcolor[rgb]{0.753,0.753,0.753}}\textbf{Device(GPU) }} & 5 SMP , 640
Cores                     \\
\hline
\end{tabular}
}
\vspace{-0.25cm}
\end{table}

\begin{table*}[!t]
\centering
\caption{Training Time (CUDA-GPU VS. Tensorflow-GPU).}
\label{table3}
\centering
{%
\renewcommand{\arraystretch}{1.2}
\begin{tabular}{|c|c|c|c|} 
\hline
\rowcolor[rgb]{0.753,0.753,0.753} {\cellcolor[rgb]{0.753,0.753,0.753}}                                                                                                                                                                  & \multicolumn{2}{c|}{\textbf{TrainingTime (second) }} & {\cellcolor[rgb]{0.753,0.753,0.753}}                                     \\ 
\hhline{|>{\arrayrulecolor[rgb]{0.753,0.753,0.753}}->{\arrayrulecolor{black}}-->{\arrayrulecolor[rgb]{0.753,0.753,0.753}}->{\arrayrulecolor{black}}|}
\rowcolor[rgb]{0.753,0.753,0.753} \multirow{-2}{*}{{\cellcolor[rgb]{0.753,0.753,0.753}}\begin{tabular}[c]{@{}>{\cellcolor[rgb]{0.753,0.753,0.753}}c@{}}\textbf{Pavia Dataset}\\\textbf{~\#Trainingsamples/\#classes~ ~~ }\end{tabular}} & \textbf{CUDA-GPU} & \textbf{Tensorflow-GPU}          & \multirow{-2}{*}{{\cellcolor[rgb]{0.753,0.753,0.753}}\textbf{Speedup }}  \\ 
\hline
200/2                                                                                                                                                                                                                                   & 0.017667          & 2.0345                           & 115.2x                                                                   \\ 
\hline
400/2                                                                                                                                                                                                                                   & 0.019695          & 2.43                             & 123.4x                                                                   \\ 
\hline
600/2                                                                                                                                                                                                                                   & 0.02487           & 3.09                             & 124.2x                                                                   \\ 
\hline
800/2                                                                                                                                                                                                                                   & 0.02797           & 4.315                            & 154.3x                                                                   \\
\hline
\end{tabular}
}
\vspace{-0.25cm}
\end{table*}

\begin{table*}[!t]
\centering
\caption{Training Time (MPI-CUDA VS. Multi-Tensorflow).}
\label{table4}
\centering
{%
\renewcommand{\arraystretch}{1.2}
\begin{tabular}{|c|c|c|c|} 
\hline
\rowcolor[rgb]{0.753,0.753,0.753} {\cellcolor[rgb]{0.753,0.753,0.753}}                                                                                                                                                                  & \multicolumn{2}{c|}{\textbf{TrainingTime (second) }} & {\cellcolor[rgb]{0.753,0.753,0.753}}                                     \\ 
\hhline{|>{\arrayrulecolor[rgb]{0.753,0.753,0.753}}->{\arrayrulecolor{black}}-->{\arrayrulecolor[rgb]{0.753,0.753,0.753}}->{\arrayrulecolor{black}}|}
\rowcolor[rgb]{0.753,0.753,0.753} \multirow{-2}{*}{{\cellcolor[rgb]{0.753,0.753,0.753}}\begin{tabular}[c]{@{}>{\cellcolor[rgb]{0.753,0.753,0.753}}c@{}}\textbf{Pavia Dataset}\\\textbf{~\#Trainingsamples/\#classes~ ~~ }\end{tabular}} & \textbf{ MPI-CUDA } & \textbf{ Multi-Tensorflow }    & \multirow{-2}{*}{{\cellcolor[rgb]{0.753,0.753,0.753}}\textbf{Speedup }}  \\ 
\hline
200/9                                                                                                                                                                                                                                   & 8.4855              & 82.762                         & 9.8x                                                                     \\ 
\hline
400/9                                                                                                                                                                                                                                   & 9.13105             & 96.72                          & 10.6x                                                                    \\ 
\hline
600/9                                                                                                                                                                                                                                   & 9.6268              & 120.32                         & 12.5x                                                                    \\ 
\hline
800/9                                                                                                                                                                                                                                   & 10.688              & 157.97                         & 14.9x                                                                    \\
\hline
\end{tabular}
}
\vspace{-0.25cm}
\end{table*}

\begin{table*}[!t]
\centering
\caption{Training Time (CUDA-GPU VS. Tensorflow-GPU).}
\label{table5}
\centering
{%
\renewcommand{\arraystretch}{1.2}
\begin{tabular}{|c|c|c|c|} 
\hline
\rowcolor[rgb]{0.753,0.753,0.753} {\cellcolor[rgb]{0.753,0.753,0.753}}                                                                                                                                                             & \multicolumn{2}{c|}{\textbf{TrainingTime (second) }} & {\cellcolor[rgb]{0.753,0.753,0.753}}                                     \\ 
\hhline{|>{\arrayrulecolor[rgb]{0.753,0.753,0.753}}->{\arrayrulecolor{black}}-->{\arrayrulecolor[rgb]{0.753,0.753,0.753}}->{\arrayrulecolor{black}}|}
\rowcolor[rgb]{0.753,0.753,0.753} \multirow{-2}{*}{{\cellcolor[rgb]{0.753,0.753,0.753}}\begin{tabular}[c]{@{}>{\cellcolor[rgb]{0.753,0.753,0.753}}c@{}}\textbf{Dataset }\\\textbf{(\#datapoint/\#features/\#classe)}\end{tabular}} & \textbf{ CUDA-GPU } & \textbf{ Tensorflow-GPU }      & \multirow{-2}{*}{{\cellcolor[rgb]{0.753,0.753,0.753}}\textbf{Speedup }}  \\ 
\hline
Iris flower (40/4/2)                                                                                                                                                                                                               & 0.018               & 1.125                          & 60.5x                                                                    \\ 
\hline
Breast Cancer (190/32/2)                                                                                                                                                                                                           & 0.0233              & 2.746                          & 117.9x                                                                   \\
\hline
\end{tabular}
}
\vspace{-0.25cm}
\end{table*}

\begin{table*}[!t]
\centering
\caption{Training Time (Tensorflow-CPU VS. Tensorflow-GPU).}
\label{table6}
\centering
{%
\renewcommand{\arraystretch}{1.2}
\begin{tabular}{|c|c|c|} 
\hline
\rowcolor[rgb]{0.753,0.753,0.753} {\cellcolor[rgb]{0.753,0.753,0.753}}                                                                                                                                                             & \multicolumn{2}{c|}{\textbf{TrainingTime (second) }}   \\ 
\hhline{|>{\arrayrulecolor[rgb]{0.753,0.753,0.753}}->{\arrayrulecolor{black}}--|}
\rowcolor[rgb]{0.753,0.753,0.753} \multirow{-2}{*}{{\cellcolor[rgb]{0.753,0.753,0.753}}\begin{tabular}[c]{@{}>{\cellcolor[rgb]{0.753,0.753,0.753}}c@{}}\textbf{Dataset }\\\textbf{(\#datapoint/\#features/\#classe)}\end{tabular}} & \textbf{ Tensorflow–CPU } & \textbf{ Tensorflow-GPU }  \\ 
\hline
Iris flower (40/4/2)                                                                                                                                                                                                               & 3.09                      & 1.125                      \\ 
\hline
Breast Cancer (190/32/2)                                                                                                                                                                                                           & 4.65                      & 2.746                      \\
\hline
\end{tabular}
}
\vspace{-0.25cm}
\end{table*}

Table \ref{table5}  shows the training times for the Iris Flower dataset with 2 classes, where CUDA-GPU achieved a speedup of 60.5X over Tensorflow-GPU. Additionally, for the Breast Cancer dataset, CUDA-GPU achieved a speedup of 117.9X. The overall results indicate that the speedup is proportionally related to the training datapoint size. Utilizing the CUDA API provides explicit control over the number of worker streaming multiprocessors (SM), thread/block, and memory management, as opposed to the implicit control in Tensorflow. Due to this explicit control, the SVM CUDA implementation demonstrates a speedup over the SVM Tensorflow implementation in the tested datasets.


Furthermore, the MPI communication overhead between nodes in the MPI-CUDA implementation may influence the overall performance, but it is only required for transferring input data at the beginning of the execution and sending the result data at the end of execution. There is no communication needed during the execution, resulting in a small overhead on system performance. Despite the MPI communication in SVM MPI-CUDA, this implementation achieved a speedup over the Multi-Tensorflow implementation in all the tested datasets.


Finally, Table \ref{table6} shows the training time for the Iris Flower dataset and the Breast Cancer dataset on Tensorflow-CPU and Tensorflow-GPU. There is no change in running the same Tensorflow implementation on both CPU and GPU, as Tensorflow has a flexible architecture that allows easy deployment over different platforms.


\section{Conclusion} \label{sec:sec5}

SVM is one of the most popular algorithms for the classification process in the ML field. Many frameworks are used to implement SVM, and in this paper, we compared two of them: CUDA and Tensorflow, on three different datasets (Pavia Centre, Iris Flower, Breast Cancer). The comparison between CUDA-GPU and Tensorflow-GPU for binary-class SVM demonstrates that the training process on CUDA-GPU achieved a speedup of 154.3X over Tensorflow-GPU for the Pavia Centre dataset with 800 sample datapoints per class and 102 features. CUDA-GPU also reached a speedup of 60.5X for the Iris Flower dataset with 40 sample datapoints per class and 4 features, and for the Breast Cancer dataset, the speedup reached 117.9X for CUDA binary implementation over Tensorflow implementation. For multi-class SVM, the MPI-CUDA implementation achieved a speedup of 14.9X over Multi-Tensorflow-GPU on the Pavia Centre dataset with 800 sample datapoints per class and 102 features, where the total number of classes is 9. Experimental results derived from three datasets show that the explicit control in CUDA API provides a speedup over the implicit control in Tensorflow. However, Tensorflow reduces development time and has the ability to be migrated to alternative hardware components.


\bibliographystyle{IEEEtran}
\bibliography{main}

\end{document}